\documentstyle[11pt,epsfig,amssymb,amsmath]{article}
\title{\bf Generalized virial theorem in Palatini $f({\mathcal{R}})$ gravity }
\author{A. S. Sefiedgar\thanks{e-mail: a-sefidgar@sbu.ac.ir}, \,K. Atazadeh\thanks{e-mail: k-atazadeh@sbu.ac.ir}
\,and H. R.  Sepangi \thanks{e-mail: hr-sepangi@sbu.ac.ir}\\
{\small Department of Physics, Shahid Beheshti university, Evin,
Tehran 19839, Iran}} \vspace{1.5cm}
\begin{document}
\maketitle
\begin{abstract}
We use the collision-free Boltzmann equation in Palatini
$f({\mathcal{R}})$ gravity to derive the virial theorem within the
context of the Palatini approach.  It is shown that the virial mass
is proportional to certain geometrical terms appearing in the
Einstein field equations which contribute to gravitational energy
and that such geometric mass can be attributed to the virial mass
discrepancy in cluster of galaxies. We then derive the velocity
dispersion relation for clusters followed by the metric tensor
components inside the cluster as well as the $f({\mathcal{R}})$
lagrangian in terms of the observational parameters. Since these
quantities may also be obtained experimentally, the
$f({\mathcal{R}})$ virial theorem is a convenient tool to test the
viability of $f({\mathcal{R}})$ theories in different models.
Finally, we discuss the limitations of our approach in the light of
the cosmological averaging used and questions that have been raised
in the literature against such averaging procedures in the context of the
present work. \noindent \vspace{5mm}\noindent\vspace{1mm}
\end{abstract}
\section{Introduction}
Dark matter is presently one of the most exciting open problems in
cosmology. There are some compelling observational evidence for the
existence of dark matter for which, the galaxy rotation curves and
mass discrepancy in cluster of galaxies are two prominent examples.
According to Newtonian gravity, galaxy rotation curves give the
velocity of matter rotating in a spiral disk as a function of the
distance from the center of galaxy according to $v(r)=\sqrt
{GM(r)/r}$. If we assume that the cluster mass obeys the relation
$M(r)=\rho\frac{4\pi r^3}{3}$, where $\rho$ is considered as a
constant density in the cluster, then the velocity increases
linearly within the cluster and drops off as the square root of $r$
outside the cluster. However, observation shows that  the velocity
remains approximately constant, that is $M\sim r$. This points to
the possible existence of a new invisible matter which is referred
to as dark matter and is distributed spherically around galaxies
\cite{1}.

What is now known as the mass discrepancy of clusters can be
understood when estimating the total mass of a cluster in two
different way; cluster masses can be deduced by summing  the
individual member masses which we shall call $M$ in total.
Alternatively, the virial theorem can be used to estimate the mass
of a cluster, $M_V$, by studying the motions of each member of the
cluster. As it turns out, $M_V$ is nearly $20-30$ times greater than
$M$ and this difference is known as the virial mass discrepancy
\cite{1}. The prime tool in dealing with the above discrepancy is to
postulate dark matter. There are several candidates for dark matter.
One possible categorization tells us that dark matter can be
baryonic or non-baryonic. The main baryonic candidates are the
Massive Compact Halo Objects (MACHO) which include brown dwarf stars
and black holes. The non-baryonic candidates are basically
elementary particles which have non-standard properties. Among the
non-baryonic candidates, we can point to axions as a solution to the
strong CP problem. However, the largest class is the Weakly
Interacting Massive Particle (WIMP) class which consists of hundreds
of, as yet, unknown particles \cite{2}. The most popular of these
WIMPs is the neutralino from supersymmetry. WIMP's interaction
cross-section with normal baryonic matter is extremely small but
non-zero, so their direct detection is a possibility. Neutrinos, may
also be considered as possible candidates for dark matter. Another
important categorization tells us that dark matter can be hot or
cold. A dark matter candidate is called hot if it was moving at
relativistic speeds at the time when galaxies were just starting to
form. It is called cold if it was moving non-relativistically at
that time. Of the above candidates only the light neutrinos would be
hot, all the others would be cold. There is, as of now, no
non-gravitational evidence for dark matter. Moreover, accelerator
and reactor experiments do not support the scenarios in which dark
matter emerges.

To deal with the question of dark matter, a great number of efforts
has been concentrated on various modifications to the Einstein field
equations \cite{3}. One such modification is that of
$f({\mathcal{R}})$ where ${\mathcal{R}}$ is the Ricci scalar.
Theories of $f({\mathcal{R}})$ modified gravity have had some
success in explaining the accelerated expansion of the universe
\cite{4,5} and account for the existence of dark matter
\cite{6,7,8}. In this paper we study $f({\mathcal{R}})$ gravity in
the context of the Palatini formalism. As is well known, starting
with the usual Einstein-Hilbert action, both the Palatini and metric
approaches result in the same field equations. However, if the
action is taken as a generic function of ${\mathcal{R}}$, then the
two approaches result in different field equations \cite{9}. Here,
we study the virial theorem within the framework described above. In
general, virial theorem plays an important role in astrophysical
objects like galaxies, clusters and super clusters. By using the
virial theorem and studying the observational data from the velocity
of each member, one can estimate the mean density of such objects,
rendering the prediction of the total mass  possible. The virial
theorem also offers interesting predictions on the stability of the
astrophysical objects. Several authors have studied the virial
theorem in models with a cosmological constant \cite{10,11}, brane
world scenarios \cite{12} and metric $f({\mathcal{R}})$ theories
\cite{13}. Our main purpose in this paper is to obtain the
generalized form of the virial theorem in $f({\mathcal{R}})$
Palatini formalism by using the collisionless Boltzmann equation. Of
course, some extra terms emerge in virial theorem which are
originated from the modified action and are geometric in nature. We
show that one may account for the virial mass discrepancy by taking
into account such extra terms. The components of the metric tensor
inside the galaxies may also be derived in terms of physical
observable quantities like the temperature of  intra cluster gas and
radius and density of the cluster core. Finding the components of
the metric tensor leads to the form of the Lagrangian in
$f({\mathcal{R}})$ gravity in terms of observable quantities. Thus,
the virial theorem is a convenient tool to test the validity of
$f({\mathcal{R}})$ models.

In what follows, we first give a brief review of the Palatini
formalism and the gravitational field equations are derived in
scalar tensor representation of $f({\mathcal{R}})$ gravity. Next, we
introduce the relativistic Boltzmann equation from which we deduce
the virial theorem with the aid of the field equations and discuss
the limitations of the assumptions made. The geometric mass and
density of a cluster is then identified in terms of the observable
quantities and the metric components are calculated inside the
cluster. We then move on to study the velocity dispersion relation
in galaxies. Finally, we present the Lagrangian which represents the
$f({\mathcal{R}})$ theory inside the cluster. Conclusions are drawn
in the last section.

\section{ Palatini $f({\mathcal{R}})$ gravity in scalar-tensor representation}
Let us start with the action
\begin{eqnarray}\label{2-1}
S_{palatini}=-\frac{1}{2\kappa}\int{d^4x\sqrt{-g}f(\mathcal{R})}+\int{d^4x\sqrt{-g}{\mathcal{L}}_{m}(g_{\mu\nu})},
\end{eqnarray}
where ${\mathcal{L}}_{m}$ is the matter lagrangian and $\kappa=8\pi
G$. The Ricci scalar is written as $\mathcal{R}$ in the context of
the Palatini formalism to point out that it is different from the
Ricci scalar, $R$, in the context of metric $f(R)$ gravity. It is
necessary to stress that ${\mathcal{R}}=g^{\mu
\nu}{\mathcal{R}}_{\mu \nu}(\Gamma)$, where ${\mathcal{R}}_{\mu
\nu}(\Gamma)$ is constructed from the connection which is
independent of the metric. In addition, in Palatini approach, the
lagrangian corresponding to matter does not depend on the
connection. Varying the action with respect to the metric and
connection, respectively, yields
\begin{eqnarray}\label{2-2}
{F(\mathcal{R})}{\mathcal{R}}_{\mu\nu} - {\frac{1}{2}}
{f({\mathcal{R}})}  g_{\mu\nu}= -\kappa {T_{\mu\nu}},
\end{eqnarray}
\begin{eqnarray}\label{2-3}
-{\overline{\nabla}}_\lambda(\sqrt{-g}{F(\mathcal{R})}g^{\mu\nu})+
{\overline{\nabla}}_\sigma(\sqrt{-g}{F(\mathcal{R})}g^{\sigma(\mu})
\delta_{\lambda}^{\nu)}=0,
\end{eqnarray}
where we have used $ \delta
{\mathcal{R}}_{\mu\nu}=\overline{\nabla}_{\lambda}\delta\Gamma_{\mu\nu}^{\lambda}-
\overline{\nabla}_{\nu}\delta\Gamma_{\mu\lambda}^{\lambda} $ with
$\overline{\nabla}_\mu$ being the covariant derivative which is
defined with the independent connection and $(\mu\nu)$ and
$[\mu\nu]$ define the symmetric and anti-symmetric parts of the
relevant parameter and $T_{\mu\nu}$ is the energy-momentum tensor.
We also denote
${F(\mathcal{R})}=\frac{d}{d{\mathcal{R}}}{f(\mathcal{R})}$. By
contracting equations (\ref{2-2}) and (\ref{2-3}), the field
equations can be written as
\begin{eqnarray}\label{2-4}
{F(\mathcal{R})}{\mathcal{R}}- 2 {f({\mathcal{R}})} =- \kappa {T},
\end{eqnarray}
\begin{eqnarray}\label{2-5}
-{\overline{\nabla}}_\lambda(\sqrt{-g}{F(\mathcal{R})}g^{\mu\nu})=0.
\end{eqnarray}
It is useful to define  a metric conformal to $g_{\mu\nu}$ as
follows
\begin{eqnarray}\label{2-6}
h_{\mu\nu}\equiv {F(\mathcal{R})}g_{\mu\nu},
\end{eqnarray}
which yields
\begin{eqnarray}\label{2-7}
\sqrt{-h}h^{\mu\nu}=\sqrt{-g} {F(\mathcal{R})}g^{\mu\nu}.
\end{eqnarray}
Now, equation (\ref{2-3}) becomes the definition of the Levi Civita
connection of $h_{\mu\nu}$ and can be solved algebraically to give
\begin{eqnarray}\label{2-8}
\Gamma^\lambda_{\mu\nu}=\frac{1}{2}
h^{\lambda\sigma}[\partial_\mu{h_{\nu\sigma}}+\partial_\nu{h_{\mu\sigma}}-\partial_\sigma{h_{\mu\nu}}].
\end{eqnarray}
By using equation (\ref{2-6}), we can write the connection in terms
of $g_{\mu\nu}$
\begin{eqnarray}\label{2-9}
\Gamma^\lambda_{\mu\nu}=\frac{g^{\lambda\sigma}}{2{F(\mathcal{R})}} [\partial_\mu{(
{F(\mathcal{R})}g_{\nu\sigma})}+\partial_\nu{(
{F(\mathcal{R})}g_{\mu\sigma})}-\partial_\sigma{(
{F(\mathcal{R})}g_{\mu\nu})}].
\end{eqnarray}
We now have an expression for $\Gamma^\lambda_{\mu\nu}$ in terms of
${\mathcal{R}}$ and $g_{\mu\nu}$ so the independent connection can
be eliminated from the field equations. In fact, using conformal
transformation in equation (\ref{2-6}), the Riemann tensor, Ricci
scalar and Einstein tensor in Palatini formalism can be derived in
terms of the metric ones \cite{14}
\begin{eqnarray}\label{2-10}
{\mathcal{R}}_{\mu\nu}=R_{\mu\nu}-\frac{3}{2}
\frac{1}{[{F(\mathcal{R})}]^2}\nabla_\mu{F(\mathcal{R})}\nabla_\nu{F(\mathcal{R})}+\frac{1}{{F(\mathcal{R})}}
(\nabla_\mu\nabla_\nu+\frac{1}{2}g_{\mu\nu}\Box){F(\mathcal{R})}.
\end{eqnarray}
Contracting equation (\ref{2-10}) by $g^{\mu\nu}$ yields the Ricci
scalar
\begin{eqnarray}\label{2-11}
{\mathcal{R}}=R-\frac{3}{2}
\frac{1}{[{F(\mathcal{R})}]^2}\nabla_\mu{F(\mathcal{R})}\nabla^\mu{F(\mathcal{R})}+\frac{3}{{F(\mathcal{R})}}
\Box{F(\mathcal{R})}.
\end{eqnarray}
Substituting equations (\ref{2-10}) and (\ref{2-11}) in (\ref{2-2}),
we may calculate the Einstein tensor
\begin{eqnarray}\label{2-12}
G_{\mu\nu}=-\frac{\kappa}{F}T_{\mu\nu}+\frac{3}{2}\frac{1}{F^2}\left[\nabla_\mu
F\nabla_\nu F-\frac{1}{2}g_{\mu\nu}\nabla_\lambda F\nabla^\lambda
F\right]-\frac{1}{F}(\nabla_\mu\nabla_\nu-g_{\mu\nu}{\Box})F-\frac{1}{2}g_{\mu\nu}\left({\mathcal{R}}-\frac{f}{F}\right).
\end{eqnarray}
As a result, the independent connections disappear and the theory is
brought to the form of GR with a modified source which only depends
on the metric and matter fields. We are now ready to introduce a
Legendre transformation $\{{\mathcal{R}},f\}\rightarrow\{\phi,V \}$
defined as
\begin{eqnarray}\label{2-13}
\phi\equiv F({\mathcal{R}}), \            \ V(\phi)\equiv
{\mathcal{R}}(\phi) F-f({\mathcal{R}}(\phi)).
\end{eqnarray}
The theory in the new representation is given by the action
\begin{eqnarray}\label{2-13/5}
S_{palatini}=-\frac{1}{2\kappa}\int{d^4x\sqrt{-g}[\phi{\mathcal{R}}-V(\phi)}]+\int{d^4x\sqrt{-g}{\mathcal{L}}_{m}(g_{\mu\nu})}.
\end{eqnarray}
Let us write the field equations in terms of the new parameters
\cite{9}
\begin{eqnarray}\label{2-14}
G_{\mu\nu}=-\frac{\kappa}{\phi}T_{\mu\nu}+\theta_{\mu\nu},
\end{eqnarray}
where
\begin{eqnarray}\label{2-15}
\theta_{\mu\nu}=\frac{3}{2{\phi}^2}(\nabla_\mu{\phi}\nabla_\nu{\phi}-\frac{1}{2}
g_{\mu\nu}\nabla^\lambda{\phi}\nabla_\lambda{\phi})
-\frac{1}{\phi}(\nabla_{\mu} \nabla_{\nu} - g_{\mu\nu}\Box)\phi -
\frac{V}{2\phi} g_{\mu\nu}.
\end{eqnarray}
It is now easy to realize that this result is the same as that of
the Brans-Dicke theory with  $\omega=\frac{3}{2}$. Introducing a
modified action, $f({\mathcal{R}})$, results in the appearance of an
effective gravitational constant. Note that
$\kappa_{eff}=\frac{\kappa}{\phi}=8\pi G_{eff}$. Therefore, we may find from
equation (\ref{2-14}) that $G_{eff}=\frac{G}{\phi}$. The
$\theta_{\mu\nu}$ tensor on the right hand side of equation
(\ref{2-14}) emerges as a new additional source for the
gravitational field.
\section{Field equations for a system of identical and collision-less point particles}
Let us now consider an isolated and spherically symmetric cluster
being described by a static and spherically symmetric metric
\begin{eqnarray}\label{3-1}
ds^2=-e^{\nu(r)}dt^2+e^{\lambda(r)}dr^2+r^2d\theta^2+r^2\sin^2\theta
d\varphi^2.
\end{eqnarray}
Suppose that the clusters are constructed from identical and
collision-less point particles (galaxies), being described by the
distribution function $f_B$. The energy-momentum tensor may be
written in terms of $f_B$ as \cite{15}
\begin{eqnarray}\label{3-2}
T_{\mu\nu}=\int{f_B m u_\mu u_\nu du},
\end{eqnarray}
where $m$ is the cluster's member mass, $u$ is the four velocity of
the galaxy and $du=\frac{du_r du_\theta du_\varphi}{u_t}$ is the
invariant volume element of the velocity space. The energy  momentum
tensor of the matter in a cluster can be represented in terms of an
effective density $\rho_{eff}$ and an effective anisotropic
pressure, with radial $p^{r}_{eff}$ and tangential $p^{\bot}_{eff}$
components \cite{10}. In other words, we have
\begin{eqnarray}\label{3-3}
\rho_{eff}=\rho\langle u_t^2\rangle ,\          \
p_{eff}^{(r)}=\rho\langle u_r^2\rangle ,  \              \
 p_{eff}^{(\bot)}=\rho\langle u_\theta^2\rangle =\rho\langle
 u_\varphi^2\rangle,
\end{eqnarray}
where $\langle\,\, \rangle$ represents the usual macroscopic
averaging.

Before going any further, a word of caution is in order at this
point. There is an interesting subtlety concerning the  energy
momentum tensor appearing on the right hand side of the Einstein
field equations defined in terms of the averaged velocities; the
fact that one can average over velocities to obtain an averaged
energy-momentum tensor does not necessarily mean that there exists a
corresponding averaged space-time metric associated to that averaged
energy-momentum tensor. This point was first raised by Flanagan
\cite{=1} and Olmo \cite{=2}. Later on, Motta and Shaw, using genal
averaging arguments \cite{=3} claimed that the averaged metric is
almost indistinguishable from that of General Relativity.
Subsequently however, it was shown in \cite{=4} that such
conclusions are not always correct because of the existence of
counter examples; infrared corrected models do not allow such
averaging of the metric. Models with high energy corrections do
admit such averaging, but it is not guaranteed in general.

The above arguments boil down to the fact that in modified gravity
theories such as  $f({\mathcal{R}})$  and within the context of the
Palatini approach, it may not be reasonable to replace the
microscopic energy-momentum tensor by its cosmological average.
Modified gravity theories imply non-linear correction terms to the
energy-momentum tensor. When applying the field equations to
cosmological scales, one may consider the microscopic structure of
all particles in the system. The existence of non-linear terms means
that one should average over all the microscopic structure of matter
and it seems as if in cosmological scales the usual macroscopic
averaging procedure is no longer valid. In Einstein general
relativity, the field equations are approximately linear on
microscopic scales. Therefore, the microscopic structure of matter
is not particularly important on macroscopic scales. The condition
in metric $f(R)$ gravity is the same as in Einstein general
relativity. But what about the Palatini formalism? The Palatini
formalism has been applied to $f({\mathcal{R}})$ gravity in the
Einstein frame \cite{=1,=2,=4} and the field equations thus obtained
are non-linear in terms of the energy-momentum tensor even on the
smallest scales. In fact, this non-linearity leads to the problem of
averaging in Palatini theories. In other words, the standard
averaging procedure may no longer be valid and results in incorrect
predictions. In addition, if one takes into account all the
microscopic structure of the matter, the palatini formalism of
$f({\mathcal{R}})$ gravity will be indistinguishable from the
standard General Relativity with a cosmological constant \cite{=3}.
Interestingly, although Palatini theories were designed to modify
gravity on large scales, they actually modify physics on the
smallest scales and leave the large scales practically unaltered. In
this work, we study  clusters containing collision-less point
particles (galaxies) within the Palatini formalism in the Jordan
frame. Here we take the cosmological averaging of the energy
momentum tensor without studying the microscopic structure of the
matter. Of course, the validity of our results should be taken in
the light of the discussion presented above.

Using $T_{\mu\nu}=(\rho+p)u_{\mu}u_{\nu}+p g_{\mu\nu}$ and
$u^{\mu}u_{\mu}=-1$ with above definitions, the field equations
become
\begin{eqnarray}\label{3-4}
e^{-\lambda}\left[\frac{\lambda'}{r}-\frac{1}{r^2}\right]+\frac{1}{r^2}=-\frac{\kappa}{\phi}\rho\langle
u_t^2\rangle
-\frac{3}{2\phi^2}(\nabla_t\phi\nabla^t\phi-\frac{1}{2}\nabla^{\lambda}\phi\nabla_{\lambda}\phi)\nonumber\\
+\frac{1}{\phi}(\nabla^t\nabla_t-\Box)\phi+\frac{V}{2\phi},
\end{eqnarray}
\begin{eqnarray}\label{3-5}
e^{-\lambda}\left[\frac{\nu'}{r}+\frac{1}{r^2}\right]-\frac{1}{r^2}=-\frac{\kappa}{\phi}\rho\langle
u_r^2\rangle
+\frac{3}{2\phi^2}(\nabla_r\phi\nabla^r\phi-\frac{1}{2}\nabla^{\lambda}\phi\nabla_{\lambda}\phi)\nonumber\\
-\frac{1}{\phi}(\nabla^r\nabla_r-\Box)\phi-\frac{V}{2\phi},
\end{eqnarray}
\begin{eqnarray}\label{3-6}
e^{-\lambda}\left[\frac{\nu'-\lambda'}{2r}-\frac{\nu'\lambda'}{4}+\frac{\nu''}{2}+\frac{{\nu'}^2}{4}\right]=-\frac{\kappa}{\phi}\rho\langle
u_\theta^2\rangle
+\frac{3}{2\phi^2}(\nabla_\theta\phi\nabla^\theta\phi-\frac{1}{2}\nabla^{\lambda}\phi\nabla_{\lambda}\phi)\nonumber\\
-\frac{1}{\phi}(\nabla^\theta\nabla_\theta-\Box)\phi-\frac{V}{2\phi},
\end{eqnarray}
\begin{eqnarray}\label{3-7}
e^{-\lambda}\left[\frac{\nu'-\lambda'}{2r}-\frac{\nu'\lambda'}{4}+\frac{\nu''}{2}+\frac{{\nu'}^2}{4}\right]=-\frac{\kappa}{\phi}\rho\langle
u_\varphi^2\rangle
+\frac{3}{2\phi^2}(\nabla_\varphi\phi\nabla^\varphi\phi-\frac{1}{2}\nabla^{\lambda}\phi\nabla_{\lambda}\phi)\nonumber\\
-\frac{1}{\phi}(\nabla^\varphi\nabla_\varphi-\Box)\phi-\frac{V}{2\phi}.
\end{eqnarray}
Another useful equation is obtained by summing the above four
equations
\begin{eqnarray}\label{3-8}
e^{-\lambda}\left[\frac{2\nu'}{r}-\frac{\nu'
\lambda'}{2}+\nu''+\frac{{\nu'}^2}{2}\right]=
-\frac{\kappa}{\phi}\rho\langle u^2\rangle
-\frac{3}{2\phi^2}(2\nabla_t\phi\nabla^t\phi)+\frac{1}{\phi}(2\nabla^t\nabla_t+\Box)\phi-\frac{V}{\phi},
\end{eqnarray}
where $\langle u^2\rangle =\langle u_t^2\rangle +\langle
u_r^2\rangle +\langle u_\theta^2\rangle +\langle u_\varphi^2\rangle
$. Since we are interested in the extra-galactic region, we assume a
small deviation from standard general relativity in which we have
$\phi=1$. Let us take $\phi=1+\epsilon g'({\mathcal{R}})$ in our
case  where $\epsilon$ is a small quantity and $g'({\mathcal{R}})$
describes the modification of the geometry due to the presence of
tensor $\theta_{\mu\nu}$ \cite{17}. Using $1/\phi\simeq1-\epsilon
g'({\mathcal{R}})$, equation (\ref{3-8}) can be written as follows
\begin{eqnarray}\label{3-9}
e^{-\lambda}\left[\frac{2\nu'}{r}-\frac{\nu'
\lambda'}{2}+\nu''+\frac{{\nu'}^2}{2}\right]=-\kappa \rho \langle
u^2\rangle -\kappa\rho_\phi,
\end{eqnarray}
where
\begin{eqnarray}\label{3-10}
-\kappa\rho_\phi\simeq \kappa\rho\langle u^2\rangle \epsilon
g'({\mathcal{R}})+\left\{
-\frac{3}{2\phi^2}[2\nabla^t\phi\nabla_t\phi]+\frac{1}{\phi}[2\nabla^t\nabla_t+\Box]\phi-\frac{V}{\phi}\right
\}\Big|_{\phi=1+\epsilon g'({\mathcal{R}})}.
\end{eqnarray}
It is not difficult to see that equation (\ref{3-10}) can be written
as
\begin{eqnarray}\label{3-11}
e^{-\lambda}\left[\frac{\nu'}{r}-\frac{\nu'
\lambda'}{4}+\frac{\nu''}{2}+\frac{{\nu'}^2}{4}\right]=-4\pi G \rho
\langle u^2\rangle -4\pi G\rho_\phi.
\end{eqnarray}
\section{The virial theorem in Palatini $f({\mathcal{R}})$ gravity}
To derive the virial theorem, we need the Boltzmann equation which
governs the evolution of the distribution function. Integrating this
equation on the velocity space when accompanied by the gravitational
field equations can yield the virial theorem. We consider an
isolated spherically symmetric cluster which is described by
equation (\ref{3-1}). The galaxies in the cluster behave like
identical, collisionless point particles. The distribution function
is denoted by $f_B$ which obeys the general relativistic Boltzmann
equation. Space-time is a time oriented Lorantzian four dimensional
manifold. The tangent bundle $T(M)$ is a real vector bundle whose
fibers at a point $x\in M$ is given by the tangent space $T_x(M)$.
The state of a particle is given by the four momentum $p\in T_x(M)$
at an event $x\in M$. The one particle phase space $P_{phase}$ is a
subset of the tangent bundle given by \cite{13,15}
\begin{eqnarray}\label{4-1}
P_{phase}:=\{(x,p)|x\in M , p\in T_x(M),p^2=-m_0^2\},
\end{eqnarray}
where $m_0$ is the particle mass. A state of a multi particle system
can be described by a continuous non-negative function $f(x,p)$. It
is defined on $P_{phase}$ and gives the number $dN$ of the particles
crossing the volume $dV$ with momenta $p$ lying within a
corresponding three-surface element $d\vec{p}$ in the momentum
space. The mean value of $f_B$ equals to the average number of
occupied particle states $(x,p)$ \cite{13,15}. Let $ \{x^\alpha\}$
with $\alpha=0\cdots 3$ be a local coordinate system in $M$, defined
in some open set $U\subset M$. Note that $\partial_t$ is timelike
future directed and $\partial_a$, $a=1,2,3$ are spacelike. Then $
\{\frac{\partial}{\partial{x^\alpha}}\}$ is the natural basis for
tangent vectors. Each tangent vector in $U$ can be written as
$p=p^\alpha\frac{\partial}{\partial x^{\alpha}}$. We can define a
system of local coordinates $\{z^A\}$, $A=0,...,7$ in $T_U(M)$ as
$z^\alpha=x^\alpha$ and $z^{\alpha+4}=p^\alpha$. A vertical vector
field over $T(M)$ is given by $\pi=p^\alpha\frac{\partial}{\partial
p^\alpha}$. The geodesic field $\sigma$, which can be constructed
over the tangent bundle, is defined as
$$\sigma=p^\alpha\frac{\partial}{\partial x^\alpha}-p^\alpha
p^\gamma \Gamma^\beta_{\alpha\gamma}\frac{\partial}{\partial
p^\beta}=p^\alpha D_\alpha,$$ where $\Gamma^\beta_{\alpha\gamma}$
are the connection coefficients. Physically, $\sigma$ describes the
phase flow for a stream of particles whose motion through spacetime
is geodesic. Therefore the transport equation for the propagation of
a particle in a curved arbitrary Riemannian spacetime is given by
the Boltzmann equation \cite{13,15}
\begin{eqnarray}\label{4-2}
\left(p^\alpha\frac{\partial}{\partial x^\alpha}-p^\alpha
p^\beta\Gamma^i_{\alpha\beta}\frac{\partial}{\partial
p^i}\right)f_B=0,
\end{eqnarray}
where $i=1,2,3$. In many applications, it is convenient to introduce
an appropriate orthonormal frame or tetrad $e^a_\mu(x)$,
$a=0\cdots3$ which varies smoothly over some coordinates in the
neighborhood of $U$ and satisfies the condition $g^{\mu\nu}e^a_\mu
e^b_\nu=\eta^{ab}$ for all $x\in U$. Any tangent vector $p^\mu$ at
$x$ can be expressed as $p^\mu=p^a e^\mu_a$, which defines the
tetrad components $p^a$. In the case of the spherically symmetric
line element given by equation (\ref{3-1}), we introduce the
following frame of orthonormal vectors \cite{13,15}
\begin{eqnarray}\label{4-3}
e^0_\mu=e^{\nu/2}\delta^0_\mu,\      \
e^1_\mu=e^{\lambda/2}\delta^1_\mu,\         \
e^2_\mu=r\delta^2_\mu,\              \
e^3_\mu=r\sin\theta\delta^3_\mu,
\end{eqnarray}
where $u^\mu$ is the four velocity of a typical galaxy, satisfying
the condition $u^\mu u_\mu=-1$ with tetrad components $u^a=u^\mu
e^a_\mu$. The relativistic boltzmann equation in tetrad components
can be written as
\begin{eqnarray}\label{4-4}
u^a e^\mu_a \frac{\partial f_B}{\partial x^\mu}+\gamma^i_{b c}u^b
u^c \frac{\partial f_B}{\partial u^i}=0,
\end{eqnarray}
where the distribution function $f_B=f_B(x^\mu,u^a)$ and
$\gamma^a_{b c}=e^a_{\mu;\nu}e^\mu_b e^\nu_c $ are the Ricci
rotation coefficients \cite{10,13,15}. We may assume that $f_B$
depends only on the radial coordinate $r$. Using equation
(\ref{2-9}), the relativistic Boltzmann equation in Palatini
formalism is obtained as
\begin{eqnarray}\label{4-5}
&u_1\frac{\partial f_B}{\partial
u_1}-\left(\frac{1}{2}u_0^2\frac{\partial \nu}{\partial
r}-\frac{u_2^2+u_3^2}{r}\right)\frac{\partial f_B}{\partial u_1}
-\frac{1}{r}u_1\left(u_2\frac{\partial f_B}{\partial
u_2}+u_3\frac{\partial f_B}{\partial u_3}\right)
-\frac{1}{r}u_3\cot\theta e^{\lambda/2}\left(u_2\frac{\partial f_B}{\partial u_3}-
u_3\frac{\partial f_B}{\partial u_2}\right)\nonumber\\
\nonumber\\
&-\frac{F'}{2F}\left[\left(u_0^2+u_1^2-u_2^2-u_3^2\right)\frac{\partial
f_B}{\partial u_1}+ 2u_1u_2\frac{\partial f_B}{\partial
u_2}+2u_1u_3\frac{\partial f_B}{\partial u_3}\right]=0.
\end{eqnarray}
Since we have assumed the system to be spherically symmetric, the
term proportional to $\cot\theta$ must be zero. Let us take
\cite{15}
\begin{eqnarray}\label{4-6}
u_0=u_t,\    \ u_1=u_r,\    \ u_2= u_\theta,\    \ u_3=u_\varphi.
\end{eqnarray}
Multiplying equation (\ref{4-5}) by $mu_rdu$, integrating over the
velocity space and assuming that $f_B$ vanishes sufficiently rapidly
as the velocities tend to $\pm\infty$, we obtain
\begin{eqnarray}\label{4-7}
\frac{\partial}{ \partial r}\left[\rho \langle u_r^2\rangle
\right]+\frac{1}{2}\frac{\partial \nu}{\partial r}\rho \left[\langle
u_t^2\rangle + 2\langle u_r^2\rangle \right]
-\frac{1}{r}\rho\left[\langle u_\theta^2\rangle +\langle
u_\varphi^2\rangle \right]+
\frac{2}{r}\rho\langle u_r^2\rangle \nonumber\\
+\frac{F'\rho}{2F}\left[\langle u_t^2\rangle +9\langle u_r^2\rangle
-\langle u_\theta^2\rangle -\langle u_\varphi^2\rangle \right]=0.
\end{eqnarray}
It is worth mentioning that in integrating equation (\ref{4-5}) we
note that the functions $F$ and $F'$ depend on the average of the
square of velocities according to equation (\ref{3-3}) via equation
(\ref{2-4}), which are assumed to be constant quantities. Now, it is
useful to multiply equation (\ref{4-7}) by $4\pi r^2$ and integrate
over the cluster volume to obtain
\begin{eqnarray}\label{4-8}
\int_0^R \rho \left[\langle u_r^2\rangle +\langle u_\theta^2\rangle
+\langle u_\varphi^2\rangle \right]4\pi r^2
dr-\frac{1}{2}\int_0^R\rho\left[\langle u_t^2\rangle +2\langle
u_r^2\rangle \right]\frac{\partial \nu}{\partial r}4\pi r^3
dr\nonumber\\-\int_0^R \frac{F'}{2F} \rho \left[\langle u_t^2\rangle
+9\langle u_r^2\rangle -\langle u_\theta^2\rangle -\langle
u_\varphi^2\rangle \right]4\pi r^3 dr=0,
\end{eqnarray}
where $R$ is the radius of the cluster.

At this point, it is appropriate to introduce some approximations.
First, consider that $\lambda'$ and $\nu'$ are small quantities.
Then the terms proportional to $\nu' \lambda'$ and $\nu'^2$ in
equation (\ref{3-11}) may be ignored. Then, assuming that
$e^{-\lambda}\approx1$ inside the cluster \cite{13}, we can write
equation (\ref{3-11}) as
\begin{eqnarray}\label{4-9}
\frac{1}{2r^2}\frac{\partial}{\partial r}\left(r^2\frac{\partial
\nu}{\partial r}\right)=-4\pi G\rho- 4\pi G \rho_\phi.
\end{eqnarray}
Second, consider that the galaxies in the cluster have velocities
much smaller than the velocity of light. In other words, $\langle
u_r^2\rangle \approx\langle u_\theta^2\rangle \approx\langle
u_\varphi^2\rangle \ll\langle u_t^2\rangle \approx1$. Now equation
(\ref{4-8}) can be written as
\begin{eqnarray}\label{4-10}
2K+\frac{1}{2}\int_0^R\rho\frac{\partial \nu}{\partial r}4\pi r^3
dr+\int_0^R \frac{F'}{2F} \rho 4\pi r^3 dr=0,
\end{eqnarray}
where
\begin{eqnarray}\label{4-11}
K=-\int_0^R \rho [\langle u_r^2\rangle +\langle u_\theta^2\rangle
+\langle u_\varphi^2\rangle ]2\pi r^2 dr,
\end{eqnarray}
is the kinetic energy of the galaxies. Multiplying equation
(\ref{4-9}) by $r^2$ and integrating yields
\begin{eqnarray}\label{4-12}
GM(r)=-\frac{1}{2} r^2 \frac{\partial \nu}{\partial r}-GM_\phi(r),
\end{eqnarray}
where we have used $M=\int_0^RdM(r)=\int_0^R4\pi \rho r^2 dr$ as the
total mass and we have also defined \\$M_\phi(r)=4\pi\int_0^r
\rho_\phi(r') r'^2 dr'$ as the geometric mass of the system. Now,
consider the definitions
\begin{eqnarray}\label{4-13}
\Omega=-\int_0^R\frac{GM(r)}{r}dM(r),
\end{eqnarray}
and
\begin{eqnarray}\label{4-14}
\Omega_\phi=\int_0^R\frac{GM_\phi(r)}{r}dM(r).
\end{eqnarray}
Multiplying equation (\ref{4-12}) by $\frac{dM(r)}{r}$ which is
equal to $\frac{4\pi \rho r^2 dr}{r}$ and integrating gives
\begin{eqnarray}\label{4-15}
\Omega=\Omega_\phi+\frac{1}{2}\int_0^R \rho \frac{\partial
\nu}{\partial r}4\pi r^3dr,
\end{eqnarray}
where $\Omega$ refers to the usual gravitational potential energy of
the system. In the end, using equation (\ref{4-10}) leads to the
generalized virial theorem in palatini formalism of
$f({\mathcal{R}})$ gravity
\begin{eqnarray}\label{4-16}
2K+\Omega-\Omega_\phi+\int_0^R \frac{F'}{2F}\rho 4\pi r^3 dr=0.
\end{eqnarray}
Recalling that $F=\phi$ in scalar-tensor representation of the
theory, the virial theorem may be written in the form
\begin{eqnarray}\label{4-17}
2K+\Omega-\Omega_\phi+\int_0^R \frac{\phi'}{2\phi}\rho 4\pi r^3
dr=0.
\end{eqnarray}
The fourth term on the left hand side of the virial equation is
originated from the relativistic Boltzmann equation in Palatini
formalism. This correction term does not exist in metric variational
approach. To represent the virial theorem in an alternative form, we
write equation (\ref{4-16}) as
\begin{eqnarray}\label{4-18}
2K-G\int_0^R \frac{M(r)dM}{r}-G\int_0^R
\frac{M_\phi(r)dM}{r}+G\int_0^R \frac{F'}{2FG}r^2\frac{dM}{r}=0,
\end{eqnarray}
or
\begin{eqnarray}\label{4-19}
2K-G\int_0^R \frac{M(r)dM}{r}-G\int_0^R
\left[M_\phi(r)-\frac{F'}{2FG}r^2\right]\frac{dM}{r}=0.
\end{eqnarray}
This equation becomes simpler by the definition
$M_\phi^{new}=M_\phi-\frac{F'}{2FG}r^2$
\begin{eqnarray}\label{4-20}
2K-G\int_0^R \frac{M(r)dM}{r}-G\int_0^R M_\phi^{new}\frac{dM}{r}=0.
\end{eqnarray}
It is convenient to introduce the radii $R_V$ and $R_\phi$
\begin{eqnarray}\label{4-21}
R_V=\frac{M^2}{\int_0^R \frac{M(r)}{r}dM(r)},
\end{eqnarray}
and
\begin{eqnarray}\label{4-22}
R_\phi=\frac{(M_\phi^{new})^2}{\int_0^R
\frac{M_\phi^{new}(r)}{r}dM(r)}.
\end{eqnarray}
In addition the virial mass $M_V$ is defined as \cite{13}
\begin{eqnarray}\label{4-23}
2K=\frac{G M M_V}{R_V}.
\end{eqnarray}
Substituting these definition in equation (\ref{4-20}) yields
\begin{eqnarray}\label{4-24}
\frac{M_V}{M}=1+\frac{(M_\phi^{new})^2 R_V}{M^2 R_\phi}.
\end{eqnarray}
For most of the observed galactic clusters, the relation $M_V/M > 3$
is true. Therefore one can easily approximate the last equation
\begin{eqnarray}\label{4-25}
\frac{M_V}{M}\approx\frac{(M_\phi^{new})^2 R_V}{M^2 R_\phi}.
\end{eqnarray}

One of the motivations of Palatini $f({\mathcal{R}})$ gravity models
is the possibility of explaining dark matter. We noted above that
some geometric terms appear in the Einstein field equations which
could effectively  play a role in gravitational energy. These
geometric terms may be attributed to a geometric mass at the
galactic or extra galactic level. They can be interpreted as dark
matter which originate from modified gravity theory. On the other
hand, dark matter provides the main mass contribution to clusters.
It means that one can ignore the mass contribution of the baryonic
mass in the clusters and estimate the total mass of the cluster by
$M_{tot}\approx M_\phi^{new}$. We also know that the virial mass is
mainly determined by the geometric mass, so that the geometric mass
could be a potential candidate for the virial mass discrepancy in
clusters. As a result, we conclude that
\begin{eqnarray}\label{4-26}
M_\phi^{new}\approx M_V\approx M_{tot}.
\end{eqnarray}
Therefore, equation (\ref{4-25}) can be written as
\begin{eqnarray}\label{4-27}
M_V\approx M \frac{R_\phi}{R_V}.
\end{eqnarray}
This shows the the virial mass is proportional to the normal
baryonic mass of the cluster, whose proportionality constant has
geometrical origins.

\section{Astrophysical applications}
\subsection{Typical values for the virial mass and radius}
Virial mass density can be written as $\rho_V=\frac{3M_V}{4\pi
R_V^3}$, where $M_V$ and $R_V$ are the virial mass and radius
respectively. Astrophysical observations together with cosmological
simulations show that the virial mass is a measure of a fixed
density such as a critical density, $\rho_c(z)$ at a special
red-shift. It means that the virial density can be represented as
$\rho_V=\delta\rho_c(z)$, where $\delta\sim200$. As is well known,
$\rho_c(z)=h^2(z)3H_0^2/8\pi G$.  The Hubble parameter is therefore
normalized to its local value,
$h^2(z)=\Omega_m(1+z)^3+\Omega_{\lambda}$, where $\Omega_m$ and
$\Omega_\lambda$ are the mass density and dark energy density
parameters respectively \cite{21}. By knowing the integrated mass of
the galaxy cluster as a function of the radius, one can estimate the
appropriate physical radius for the mass measurement. The commonly
used radii are $r_{200}$ or $r_{500}$. These radii lie within the
radii corresponding to the mean gravitational mass density of the
matter $\langle \rho_{tot}\rangle =200\rho_c$ or $500\rho_c$. A
useful radius is $r_{200}$ to find the virial mass. By studying the
values of $r_{200}$ for various clusters, one can deduce that a
typical value for $r_{200}$ is $2Mpc$ approximately. The
corresponding masses for these radii are defined as $M_{200}$ and
$M_{500}$. In general, $M_V=M_{200}$ and $R_V=r_{200}$ are assumed
as a virial mass and radius \cite{24}.

\subsection{Geometric mass and geometric radius from galactic cluster observations}
Intra cluster gas has an important contribution to the baryonic mass
of the clusters of galaxies. The following equation provides a
reasonably good description of the observational date  \cite{13,24}
\begin{eqnarray}\label{5-1}
\rho_g=\rho_0 \left(1+\frac{r^2}{r_c^2}\right)^{-\frac{3\beta}{2}},
\end{eqnarray}
where $r_c$ is the core radius and $\rho_0$ and $\beta$ are
constants. If we assume that the observed X ray emission from the
hot, ionized intra cluster gas is in isothermal equilibrium, then
the pressure $p_g$ of the gas satisfies the equation of state
$p_g=\left(\frac{k_BT_g}{\mu m_g}\right)\rho_g$, where $k_B$ is the
Boltzmann constant, $T_g$ is the gas temperature, $\mu\approx0.61$
is the mean atomic weight of the particles in the cluster gas and
$m_p$ is the proton mass \cite{13,24}. Using Jeans equation
\cite{1}, the total mass distribution can be obtained as
\cite{12,13,24}
\begin{eqnarray}\label{5-2}
M_{tot}(r)=-\frac{k_B T_g}{\mu m_pG}r^2\frac{d}{dr}\ln{\rho_g}.
\end{eqnarray}
Substitution of the mass density of the cluster gas in equation
(\ref{5-1}) gives the total mass inside the cluster \cite{13,24}
\begin{eqnarray}\label{5-3}
M_{tot}(r)=\frac{3k_B\beta T_g}{\mu m_p G}\frac{r^3}{r_c^2+r^2}.
\end{eqnarray}
Now, $f({\mathcal{R}})$ theory tells us that the total mass of the
cluster is $M_{tot}(r)=4\pi \int_0^r (\rho_g+\rho_\phi)r^2 dr$,
where it is assumed that the main part of the baryonic mass of the
cluster is in the form of intra cluster gas. Therefore, $M_{tot}(r)$
satisfies the relation
\begin{eqnarray}\label{5-4}
\frac{dM_{tot}(r)}{dr}=4\pi r^2 \rho_g(r) + 4\pi r^2 \rho_\phi(r).
\end{eqnarray}
Since we have estimated the quantities $M_{tot}(r)$ and $\rho_g$,
the expression for the geometric mass density can be readily
obtained
\begin{eqnarray}\label{5-5}
4\pi \rho_\phi(r)=\frac{3K_B\beta T_g\left(r^2+3r_c^2\right)}{\mu
m_p\left(r_c^2+r^2\right)^2}- \frac{4\pi
G\rho_0}{\left(1+\frac{r^2}{r_c^2}\right)^{\frac{3\beta}{2}}}.
\end{eqnarray}
In the limit $r\gg r_c$, the geometric density takes the simple form
\begin{eqnarray}\label{5-6}
4\pi \rho_\phi(r)\approx \left[\frac{3K_B\beta T_g}{\mu m_p}-{4\pi
G\rho_0}r_c^{3\beta} r^{2-3\beta}\right]\frac{1}{r^2}.
\end{eqnarray}
Also, using equation (\ref{5-5}), we can easily write the geometric mass as
\begin{eqnarray}\label{5-7}
GM_{\phi}(r)=4\pi \int_0^r r^2 \rho_\phi(r)dr=\frac{3K_B\beta
T_g}{\mu m_p}\frac{r}{1+\frac{r_c^2}{r^2}}-4\pi G\rho_0 \int_0^r
\frac{r^2 dr}{(1+\frac{r^2}{r_c^2})^{\frac{3\beta}{2}}}.
\end{eqnarray}
In the limit $r\gg r_c$, the geometric mass takes the simple form
\begin{eqnarray}\label{5-8}
GM_{\phi}(r)\approx \left[\frac{3K_B\beta T_g}{\mu m_p}-\frac{4\pi G
\rho_0 r_c^{3\beta}r^{2-3\beta}}{3(1-\beta)}\right]r.
\end{eqnarray}

Observations show that the intra cluster gas has a small
contribution to the total mass \cite{13,18,24,21,23}. This means
that the gas density and  mass contributions can be neglected
compared to the geometric density and mass, hence
\begin{eqnarray}\label{5-9}
4\pi G\rho_\phi(r)\approx \left(\frac{3K_B\beta T_g}{\mu
m_p}\right)r^{-2},
\end{eqnarray}
\begin{eqnarray}\label{5-10}
GM_{\phi}(r)\approx \left(\frac{3K_B\beta T_g}{\mu m_p}\right)r.
\end{eqnarray}
Now, it is easy to derive the metric tensor component $e^\nu$ inside
the cluster. From equation (\ref{4-9}), we deduce that
$r^2\nu'=-2GM_{tot}(r)\approx-2GM_\phi(r)$, which leads to
\begin{eqnarray}\label{5-11}
e^\nu\approx C_\nu r^{2s},
\end{eqnarray}
where $C_\nu$ is the integration constant and $s$ is defined as
\begin{eqnarray}\label{5-12}
s=-\frac{3k_B\beta T_g}{\mu  m_p}.
\end{eqnarray}
The other coefficient in the metric tensor, $e^{-\lambda}$, can be
estimated approximately as
\begin{eqnarray}
e^{-\lambda}\approx1-\frac{2GM_{tot}}{r},
\end{eqnarray}
which is the standard expression which, assuming $M_{tot}\approx
M_\phi$,  may be written as $e^{-\lambda}\approx 1-\frac{6k_B\beta
T_g}{\mu m_p}=1+2s$. Of course, there is no guarantee as to the
accuracy of this approximation inside the cluster. As a result,
astrophysical observations can be helpful in deriving the metric
components inside the cluster in the context of $f({\mathcal{R}})$
theory . However, how can we estimate the upper bound for the cut
off of the geometric mass? This happens in a special point where the
decaying density profile of the geometric density associated with
the cluster becomes smaller than the average energy density of the
universe. We assume that the two densities become equal at the
radius $R_{\phi} ^{cr}$. In other words, we can write
$\rho_\phi(R_\phi ^{cr})=\rho_{universe}$. Let us set
$\rho_{universe}=\rho_c=\frac{3H^2}{8\pi G}=4.6975 \times
10^{-30}h^2_{50} g/{cm^{-3}}$ where $H=50h_{50}km/{Mpc}/s$
\cite{13,24}. By using equation (\ref{5-9}), we obtain
\begin{eqnarray}\label{5-13}
R_\phi^{(cr)}=\left(\frac{3k_B\beta T_g}{\mu m_p G
\rho_c}\right)^{\frac{1}{2}}=91.33\sqrt{\beta}\left(\frac{k_B
T_g}{5keV}\right)^{\frac{1}{2}}h_{50}^{-1}Mpc.
\end{eqnarray}
We may also derive the total geometric mass in equation (\ref{5-10}) corresponding to this
radius as
\begin{eqnarray}\label{5-14}
M_\phi^{(cr)}=M_\phi({R^{(cr)}_\phi})=4.83\times10^{16}\beta^{3/2}\left(\frac{k_B
T_g}{5keV}\right)^{\frac{3}{2}}h_{50}^{-1}M_\odot.
\end{eqnarray}
This value is consistent with the mass distribution observations in
clusters. Within $f({\mathcal{R}})$ gravity, geometric mass effects
are beyond the virial radius, which is about a few $Mpc$.

\subsection{Radial velocity dispersion in galactic clusters}
We can write the virial mass in terms of the characteristic velocity dispersion $\sigma_1$ as \cite{23}
\begin{eqnarray}\label{6-1}
M_V=\frac{3}{G}\sigma_1^2 R_V,
\end{eqnarray}
where $3\sigma_1^2=\sigma_r^2$. Let us consider an isotropic
velocity dispersion. Then, we have $\langle u^2\rangle =\langle
u_r^2\rangle +\langle u_\theta^2\rangle +\langle u_\varphi^2\rangle
=3\langle u_r^2\rangle =3\sigma_r ^2$ where $\sigma_r^2$ is the
radial velocity dispersion. Radial velocity dispersion relation for
clusters in $f({\mathcal{R}})$ gravity can be obtained from equation
(\ref{4-7}) as
\begin{eqnarray}\label{6-2}
\frac{d}{d r}(\rho \sigma_r^2)+\frac{1}{2}\rho \frac{d \nu}{d
r}+\frac{F'}{2F}\rho=0,
\end{eqnarray}
from which one can deduce a relation for $\nu'$. Also  equation
(\ref{4-9}) yields a relation for $\nu'$
\begin{eqnarray}\label{6-3}
\nu'=-\frac{1}{r^2}[2GM_\phi(r)+2GM(r)+2c],
\end{eqnarray}
where $c$ is an integration constant. By eliminating $\nu'$ from the
last two equations, we obtain
\begin{eqnarray}\label{6-4}
\frac{d}{dr}(\rho
\sigma_r^2)=-\frac{F'}{2F}\rho+\frac{1}{r^2}[GM_\phi(r)+GM(r)+c]\rho.
\end{eqnarray}
Integration now gives the solution as follows
\begin{eqnarray}\label{6-5}
\sigma_r^2(r)=-\frac{1}{\rho}\int\frac{F'}{2F}\rho
dr+\frac{1}{\rho}\int[GM_\phi(r)+GM(r)+c]\frac{\rho
dr}{r^2}+\frac{c'}{\rho}.
\end{eqnarray}
We can apply equation (\ref{6-5}) to a special case where the
density $\rho$ is chosen as \cite{13}
\begin{eqnarray}\label{6-6}
\rho(r)=\rho_0 r^{-\gamma},
\end{eqnarray}
with $\rho_0$ and $\gamma\neq1,3$ being positive constants. As $\rho$
is the normal matter density inside the cluster, it yields the normal
matter mass profile $M(r)=4\pi \rho_0r^{3-\gamma}/(3-\gamma)$.
According to equation (\ref{5-10}), the geometric mass
$GM_\phi(r)\approx q_0r$, where $q_0=3k_B\beta T_g/{\mu m_p}$. The
radial velocity  dispersion for $\gamma\neq 1,3$ will be
\begin{eqnarray}\label{6-7}
\sigma_r^2(r)=-r^\gamma \int\frac{F'}{2F}r^{-\gamma}
dr-\frac{q_0}{\gamma}-\frac{2\pi G
\rho_0}{(\gamma-1)(3-\gamma)}r^{2-\gamma}-\frac{c}{\gamma+1}\frac{1}{r}+\frac{c'}{\rho_0}r^\gamma.
\end{eqnarray}
For $\gamma=1$ we find
\begin{eqnarray}\label{6-8}
\sigma_r^2(r)=-r \int\frac{F'}{2F}r^{-1} dr-q_0+2\pi G\rho_0r\ln
r-\frac{c}{2r}+\frac{c'}{\rho_0}r,
\end{eqnarray}
and for $\gamma=3$ we have
\begin{eqnarray}\label{6-9}
\sigma_r^2(r)=-r^3 \int\frac{F'}{2F}r^{-3} dr-\frac{q_0}{3}+\pi
G\rho_0\left(\ln
r+\frac{1}{4}\right)-\frac{c}{4}\frac{1}{r}+\frac{c'}{\rho_0}r^3.
\end{eqnarray}
The observed data can usually be translated to a specific function
for velocity dispersion relation. Then one can compare the observed
velocity dispersion with  prediction in modified $f({\mathcal{R}})$
gravity to compare the different theoretical scenarios.

\section{The Lagrangian}
We recognized the possibility of finding the metric tensor
components in $f({\mathcal{R}})$ gravity using the virial theorem in
the previous sections. One of the other consequences is to find the
lagrangian $f({\mathcal{R}})$ of the theory, in other words, the
form that $f({\mathcal{R}})$ can take. We start from the field
equations in the standard representation. Equations (\ref{3-4}) and
(\ref{3-5}) then yield
\begin{eqnarray}\label{7-1}
F''-\frac{\nu'+\lambda'}{2}F'+\frac{\nu'+\lambda'}{r}F-\frac{3}{2}\frac{F'^2}{F}=0.
\end{eqnarray}
Also, from Equations (\ref{3-6}) and (\ref{3-7}) we have
\begin{eqnarray}\label{7-2}
-\frac{\nu'
\lambda'}{4}+\frac{\nu''}{2}+\frac{\nu'^2}{4}+\frac{1}{r^2}(e^\lambda-1)=
-\frac{3}{4}\left(\frac{F'}{F}\right)^2+\frac{1}{2}\frac{F''}{F}
+\left(\frac{\nu'-\lambda'}{4}-\frac{1}{r}\right)\frac{F'}{F}.
\end{eqnarray}
Equation (\ref{2-2}) with $\mu=\nu=0$, using equation (\ref{2-10})
leads to
\begin{eqnarray}\label{7-3}
f=Fe^{-\lambda}\left[\frac{\nu'
\lambda'}{2}-\nu''-\frac{\nu'^2}{2}-\frac{2\nu'}{r}+\left(\frac{3\nu'-\lambda'}{2}+\frac{2}{r}\right)\frac{F'}{F}+
\frac{F''}{F}\right].
\end{eqnarray}
Now, we can write the Ricci scalar from equations (\ref{2-4}) and
(\ref{2-11}) as follows
\begin{eqnarray}\label{7-4}
R=\frac{2f}{F}-3e^{-\lambda}\left[-\frac{1}{2}\left(\frac{F'}{F}\right)^2+\left(\frac{F''}{F}\right)
+\left(\frac{\nu'-\lambda'}{2}+\frac{2}{r}\right)\left(\frac{F'}{F}\right)\right].
\end{eqnarray}
Here, $R$ is derived in the context of metric formalism. In fact, we
have replaced all the parameters in terms of those appearing in the
metric formalism which would then lead to the metric Ricci scalar.
In the derivation of the above equations, we have neglected the
contribution from the baryonic matter. Equation (\ref{5-11})
together with the assumption $e^\lambda=$const. inside the clusters
then become reasonable to use. Now, equation (\ref{7-1}) can be
written as
\begin{eqnarray}\label{7-5}
FF''-\frac{s}{r}F'F-\frac{3}{2}F'^2+\frac{2s}{r^2}F^2=0,
\end{eqnarray}
for which a solution is
\begin{eqnarray}\label{7-6}
F=\frac{4(s^2+6s+1)r^{(-s-1-\sqrt{s^2+6s+1})}}{c},
\end{eqnarray}
where $c$ is an integration constant and we set $c=1$. The component
$e^\lambda$ can be derived from Equation (\ref{7-2}) by using the
expression for $F$
\begin{eqnarray}\label{7-7}
e^\lambda=1+s-s^2+\frac{(s+1+\sqrt{s^2+6s+1})}{4}\left(5-3s-\sqrt{s^2+6s+1}\right).
\end{eqnarray}
Now, equation (\ref{7-3}) gives
\begin{eqnarray}\label{7-8}
\frac{f}{F}=\frac{e^{-\lambda}}{r^2}\left[-2s-2s^2+(s+1+\sqrt{s^2+6s+1})(-2s+\sqrt{s^2+6s+1})\right],
\end{eqnarray}
and equation (\ref{7-4}) results in
\begin{eqnarray}\label{7-9}
R=\frac{e^{-\lambda}}{r^2}\left[\frac{(s+1+\sqrt{s^2+6s+1})(-5s+3+\sqrt{s^2+6s+1})}{2}-4s-4s^2\right].
\end{eqnarray}
Recall that ${\mathcal{R}}$ is the Ricci scalar in Palatini
formalism whereas $R$ is the Ricci scalar in metric formalism.
Finally, we can easily obtain $f$ as a function of $R$ as follows
\begin{eqnarray}\label{7-10}
f=f_0R^{\frac{s+3+\sqrt{s^2+6s+1}}{2}},
\end{eqnarray}
where
\begin{eqnarray}\label{7-11}
f_0=\frac{4(s^2+6s+1)\left[-2s-2s^2+(s+1+\sqrt{s^2+6s+1})(-2s+\sqrt{s^2+6s+1})\right]}
{\left[\frac{(s+1+\sqrt{s^2+6s+1})(-5s+3+\sqrt{s^2+6s+1})}{2}-4s-4s^2\right]^{\frac{s+3+\sqrt{s^2+6s+1}}{2}}}
e^{{\lambda\frac{(s+1+\sqrt{s^2+6s+1})}{2}}}.
\end{eqnarray}
One may also obtain the action in terms of the Ricci scalar in
Palatini formalism. In either case, knowing the physical parameters
such as  gas density and temperature, one can deduce the action for
modified gravity.
\section{Conclusions}
Virial theorem is a convenient tool with which to derive the mean
density of galaxy clusters and it can therefore predict the total
mass of clusters. We have used the relativistic Boltzmann equation
within the context of Palatini $f({\mathcal{R}})$ theory to derive
the virial theorem. To write the field equations, we have applied
the standard cosmological averaging for the energy-momentum tensor.
The virial mass is mainly determined by the geometric mass which is
associated to the geometrical terms in the gravitational field
equations. The generalized virial mass implicitly includes the
effects of dark matter and may therefore be used to describe the
dynamics of clusters. If equations (\ref{4-27}) and (\ref{5-13})
accompany the assumption $R_\phi\approx R_\phi^{(cr)}$, one can
estimate the virial mass as $M_V\approx
91.33\sqrt{\beta}(\frac{k_BT_g}{5keV})^{\frac{1}{2}}h_{50}^{-1}\frac{M}{R_V(Mpc)}$
which, by having the physical parameters of a cluster, leads to a
specific value. In fact the virial mass can be approximated to give
the total mass of clusters. In this paper, we have found that the
virial mass or the total mass of  clusters can also be obtained from
the velocity dispersion relations. However, uncertainties in
observational data suppresses the exact value of the virial mass. It
seems that the gravitational lensing of light in $f({\mathcal{R}})$
theories gives more exact values for the total mass of  clusters. In
any case, since the virial theorem results can be compared with
observational data, one can apply it to different theories to test
their viability. We also derived, in effect, the Lagrangian of the
theory in terms of the metric Ricci scalar. Finally, we discussed
the limitations of our approach which is rooted on the assumption
that there exists an averaged metric which is the outcome of putting
on the right hand side of the field equations an averaged energy
momentum tensor.\noindent\vspace{5mm}\\
{\bf Acknowledgement}\noindent\vspace{2mm}\\
We would like to thank the anonymous referee for invaluable comments
and criticisms.


\end{document}